\documentstyle[prl,aps]{revtex}
\def\be{\begin{equation}}
\def\ee{\end{equation}}
\def\bea{\begin{eqnarray}}
\def\eea{\end{eqnarray}}

\begin{document}

\draft

 \preprint{}

\title
{Destruction of bulk ordering by surface randomness}
\author
{D.E. Feldman$^{1,2}$ and V.M. Vinokur$^1$}
\address
{$^1$Materials Science Division, Argonne National Laboratory,
9700 South Cass, Argonne, Illinois 60439\\
$^2$Landau Institute for Theoretical Physics, Chernogolovka,
Moscow region 142432 Russia} \maketitle
\begin{abstract}
We demonstrate that the arbitrarily weak quenched disorder on the
surface of a system of continuous symmetry destroys long range
order in the bulk, and, instead, quasi-long range order emerges.
Correlation functions are calculated exactly for the  two- and
three-dimensional $XY$ model with surface randomness via the
functional renormalization group. Even at strong quenched disorder
the three-dimensional $XY$ model possesses topological order. We
also determine roughness of a domain wall in the presence of
surface disorder.
\end{abstract}

The arbitrarily weak quenched disorder in the bulk of a system of
continuous symmetry destroys long range order inherent to the
pure system \cite{Larkin} provided that disorder breaks not only
the translational symmetry but also the symmetry with respect to
transformations of the order parameter, as e.g. random anisotropy
in amorphous magnets does. This fundamental fact governs all the
physics of condensed matter and results in a wealth of observed
static and dynamics behaviors of real solids.

In many cases noticeable disorder presents only at the surface.
Not surprisingly, surface randomness modifies the critical
behavior near the surface \cite{HankeKardar}, yet the common
expectation is for the bulk properties to remain intact. In this
Letter we show that arbitrarily weak surface disorder destroys
long range order in the bulk of a system of continuous symmetry
at the arbitrarily low temperature.

The predicted effect occurs in a rich variety of systems. Examples
include crystal ordering in solids grown on a disordered
substrate, liquid crystals interacting with an inhomogeneous
surface, superconducting vortices pinned by surface impurities,
etc. There are also many two-dimensional systems with edge
randomness, e.g. superconducting films with columnar defects in a
part of the film or films with a rough edge \cite{Kes}.


The reason as to why surface impurities, however weak, break long
range bulk order is that the bulk contributes little to the energy
of long-wave Goldstone modes: the surface random energy of
long-wave excitations turns out to be greater than the
corresponding bulk energy. As a result, the inhomogeneous state
becomes favorable energetically. Note that ordering survives in
the regions of the size less than the distance of these regions
from the surface. In other words, if the distance between the two
points is greater than their separation from the surface, the
order parameter is different in those points. While long range
order breaks down, topological order survives and quasi-long range
order emerges. This means that the correlation length is infinite
and that the correlation functions obey a slow logarithmic
dependence of the distance.

An easy way to understand the main result of the Letter is based
on Imry-Ma arguments \cite{Larkin,ImryMa}. Let us consider a
region of size $L$ near the surface and compare the energies of
ordered and disordered states of the region. If long range order
is broken on the scales of the order of $L$ the loss in the bulk
(elastic) energy is $E_{\rm bulk}\sim L^D/L^2$, where $D$ is the
space dimension. The energy gain from the interaction with surface
impurities scales as $E_{\rm surface}\sim L^{(D-1)/2}$. If $D<3$
then $E_{\rm surface}>E_{\rm bulk}$. Hence, the disordered state
is favorable at $D<3$. On the other hand, long range order is
favorable at $D>3$. The case $D=3$ is marginal and requires more
quantitative consideration. We will see that in this case long
range order is absent for the arbitrarily weak disorder similar to
$D<3$.

We begin with the analysis of the exactly solvable Larkin model
\cite{Larkin}. It is defined by the Hamiltonian

\be \label{1} H=\frac{J}{2}\int(\nabla\phi)^2d^2xdz-\int
h(x)\phi(x,z=0) d^2x \ee where $\phi$ is the order parameter, $h$
the random field, the $z$-axis is perpendicular to the disordered
surface. At zero temperature we search for the energy minimum. To
find the field configuration $\phi^*(x,z)$ at zero temperature we
calculate the variation of the Hamiltonian (\ref{1}) and make the
Fourier transform with respect to the $x$ coordinates:

\be \label{2} J(q_{||}^2-\partial_z^2)\phi^*(q_{||},z)= h(q_{||})
\delta(z) \ee Substituting the solution of this equation \be
\label{sol} \phi^*(q_{||},z)= \phi^*(q_{||},0)\exp(-|q_{||}|z) \ee
into Eq. (\ref{1}) we get the energy

\be \label{3} H^*=\int \frac{d^2
q_{||}}{(2\pi)^2}\big[|q_{||}||\phi^*(q_{||},0)|^2
-h(q_{||})\phi^*(-q_{||},0)\big].
 \ee

From Eq. (\ref{3}) we find
$\phi^*=\exp(-|q_{||}|z)h_{q_{||}}/(J|q_{||}|)$. Assuming that the
random field $h$ is Gaussian and $\langle h(q)h(p)\rangle=\Delta
\delta (p+q)$ one finds the correlation function $G(r_1,r_2)=
\langle(\phi(r_1)-\phi(r_2))^2\rangle$. For example, if
$z_1=z_2\ll(r_1-r_2)$ then \be \label{G} G=\Delta/(\pi
J^2)\ln(|x_1-x_2|/z_1)\ee If $x_1=x_2$, $z_1\ll z_2 $ then
$G=\Delta/(2\pi J^2)\ln(z_2/z_1)$. Since the correlation function
is unlimited, there is no long range order. We will see that the
same behavior is present in more complicated systems too, e.g. in
the 3D $XY$ model with surface disorder.

As the first nontrivial example we consider a domain wall in a
media with disordered surface \cite{RandomManifolds}. The domain
wall is rough in the presence of bulk disorder
\cite{RandomManifolds}. We show that surface disorder also makes
it rough. The shape of the domain wall is described by the
displacement of the wall as a function of $D-1$ coordinates
$y=\phi({\bf x},z)$ where ${\bf x}$ and $z$ are the coordinates in
a plane perpendicular to the surface, $z$ being the direction
perpendicular to the surface. The Hamiltonian differs from
(\ref{1}) only by the dependence of the random contribution on
$\phi$: in the case of the random-bond disorder
\cite{RandomManifolds} it is a Gaussian $\delta$-correlated random
variable $V({\bf x}, \phi)$, $\langle
V(x_1,\phi_1)V(x_2,\phi_2)\rangle\sim\delta(x_1-x_2)\delta(\phi_1-\phi_2)$.
 We will demonstrate that the $\langle\phi\phi\rangle$ correlation
function exhibits a power-law distance dependence:

     \be \label{cor}
          \langle(\phi(r_1)-\phi(r_2))^2\rangle\sim|r_1-r_2|^{2\zeta}.
          \ee

Introducing the field $\phi^ *(q_{||},z)$ as above and repeating
the derivation of Eq. (\ref{3}) we get

\be \label{4} H^*=\int \frac{d^{D-2}
q_{||}}{(2\pi)^{D-2}}\frac{|q_{||}||\phi^*(q_{||},0)|^2}{2} +\int
d^{D-2}x V(x,\phi^*(x,0)). \ee After replica averaging over
disorder we get the effective replica Hamiltonian

\be \label{5} H_R=\int \frac{d^{D-2}
q_{||}}{(2\pi)^{D-2}}\sum_a\frac{|q_{||}||\phi_a^*(q_{||},0)|^2}{2T}
-\int d^{D-2}x \sum_{ab} \frac{R(\phi_a^*(x)-\phi_b^*(x))}{2T^2},
\ee where $a$ and $b$ are replica indices, $R$ is the function
describing disorder. The problem can be studied with the
functional renormalization group (RG)  following the line of Ref.
\cite{Fisher1}. At each RG step we integrate out the momenta from
the interval $1/(as)<q_{||}<1/a$, where $a$ is the ultraviolet
cutoff and $s>1$, and make the rescaling $q_{||}\rightarrow
q_{||}/s$, $x\rightarrow sx$, $\phi^*\rightarrow s^\zeta\phi^*$.
The first term of Eq. (\ref{5}) depends on $q_{||}$
nonanalytically and hence does not renormalize \cite{ZJ}. Thus,
the temperature obeys the RG equation $dT/d\ln L=(3-D-2\zeta)T $
and we find a zero-temperature fixed point. Power counting shows
that the whole function $R(\phi)$ is a relevant operator near
$D=4$ in this fixed point. The RG equation for $R$ can be derived
exactly in the same way as in Ref. \cite{Fisher1} and to the first
order in $\epsilon=4-D$ has the same form as in Ref.
\cite{Fisher1}

\be \label{6} dR(\phi)/d\ln L = (\epsilon-4\zeta)R(\phi)+\zeta\phi
R'(\phi)+R''(\phi)^2/2-R''(\phi)R''(\phi=0)\ee where the factor
$S_2/(2\pi)^2=1/(2\pi)$ is absorbed into $R$. The solution of this
equation can be read out of Ref. \cite{Fisher1} and gives us
$\zeta=0.2083\epsilon$ in the correlation function (\ref{cor}).
Using the result (\ref{sol}) for the displacement field $\phi$ one
gets the same behavior
$\langle(\phi(x_1,z_1)-\phi(x_2,z_2))^2\rangle\sim|x_1-x_2|^{2\zeta}$
in the bulk as at the surface provided that $|x_1-x_2|>z_1\sim
z_2$. Thus, the domain wall is rough in 3 dimensions in the
presence of surface disorder.

Next we consider the $XY$ model with surface randomness. It is
described by the same Hamiltonian (\ref{5}) as the domain wall
problem. However, due to the periodicity of the $XY$ model the
function $R(\phi)$ has period $2\pi$. One can apply this model for
the description of vortices in a superconductor with disordered
surface \cite{vortices}. The field $\phi$ describes displacements
of the vortices from the regular positions in the Abrikosov
lattice. Random potential is associated with impurities which pin
vortices at the surface.

Due to the periodicity of the Hamiltonian there is no
renormalization of the order parameter, i.e. $\zeta=0$. The RG
equations have exactly the same structure as in the domain wall
problem in dimension $D+1$. Since the dimension 4 is the critical
dimension in the domain wall problem, the random $XY$ model has
the critical dimension $D=3$. Thus, we expect the logarithmic
situation in 3 dimensions: the theory is asimptotically free and
$R(\phi)$ can be represented as $R(\phi, L)=R^*(\phi)/\ln L$. The
${\it exact}$ RG equation for $R^*$ has the same structure as the
one-loop RG equation (\ref{6}) at $\epsilon=1$ and $\zeta=0$. The
analytical solution of this equation is known
\cite{vortices,review} and gives the following exact result for
the correlation function:
$$\langle(\phi(x_1,z_1)-\phi(x_2,z_2))^2\rangle=\frac{\pi^2}{9}\ln\ln
|x_1-x_2|$$ at $|x_1-x_2|>z_1,z_2$.
 The possibility to obtain
an exact result makes the 3D $XY$ model with surface disorder a
good candidate for a quantitative test of the functional RG.
Since the correlation function is unlimited at large $r_1-r_2$,
long range order is absent. However, the correlation function
depends on the distance very slow. Such a  situation can be
described as super-quasi-long range order \cite{review}.

Our treatment of the $XY$ model in terms of the elastic
Hamiltonian (\ref{5}) is justified in the absence of topological
defects. Similar to Ref. \cite{review} we do not expect that the
defects are relevant at weak disorder. At strong disorder,
dislocations also do not proliferate deep in the bulk. Indeed, let
us consider a dislocation of size $L$. The distance $R$ of any of
its points from the surface is less than the dislocation length
$L$ since otherwise the dislocation would not interact with the
surface and should be energetically unfavorable. We estimate the
energy gain associated with the disorder as $E_{\rm disorder}\sim
\sqrt\Delta\sqrt{LR}$, where $\Delta$ is the average square of the
random field, and the bulk energy loss as $E_{\rm bulk}\sim J L\ln
R$. One sees that the bulk loss is greater than the surface gain
for large $L$ even at strong disorder $\sqrt\Delta\gg J$. Thus,
proliferation of large dislocation loops is unfavorable and the
system renders topological order.

In two dimensions the $XY$ model has only quasi-long range order
at low temperatures even in the absence of disorder.  Edge
disorder modifies correlation functions but does not destroy
quasi-long range order as we show below.

A recent experimental realization of a related two-dimensional
system with edge randomness is a superconducting film with
columnar defects in a part of it \cite{Kes}. Vortices pinned by
defects create a random potential near the boundary between the
pure and impure parts of the film. This potential affects the
vortex lattice in the pure part. An anisotropic film where
vortices are free to move only in one direction can be described
by the $XY$ model with edge disorder \cite{vortices}.

One cannot use the $3-\epsilon$ expansion in this problem since no
zero temperature fixed point exists in 2 dimensions: the scaling
dimension of the temperature $\Delta_T=D-2$ becomes zero in two
dimensions. Physically this result is related to the fact that
thermal fluctuations have qualitatively the same effect on the
ordering in the system as impurities: both destroy long-range
order. Hence, we have to develop an RG procedure directly in two
dimensions at nonzero temperature.

Similar to the derivation of Eq. (\ref{5}) we obtain the following
one-dimensional replica Hamiltonian for the field $\phi^*$ at the
edge (in the absence of vortices)

\be \label{7} H_R=\int \frac{d
q_{||}}{(2\pi)}\sum_a\frac{J|q_{||}||\phi_a^*(q_{||},0)^2|}{2T}
-\int dx \sum_{ab} \sum_k\frac{\Delta_k\cos k
(\phi_a^*-\phi_b^*)}{T^2} \ee where $k=nk_{\rm min}$, $k_{\rm
min}$ being the smallest value of $k$ allowed by the symmetry, $n$
an integer.
 Note that in this expression
$\phi^*$ is not a zero-temperature field configuration in contrast
to the previous problems. The field $\phi$ in the bulk of our 2D
system includes two contributions: one is related to $\phi^*$ via
Eq. (\ref{sol}) and the other is a free thermally fluctuating
field.

We first neglect vortices and then check how they change the
behavior of the system. At $T>T_c=\pi J/k^2_{\rm min}$ all the
nonlinear terms in $H_R$ are irrelevant. At high temperatures the
correlation function is the same as in the absence of disorder and
has the form $\langle(\phi^*(x_1)-\phi^*(x_2))^2\rangle=2T/(\pi
J)\ln |x_1-x_2|$. Below $T_c$ the cosine term with the minimal $k$
becomes relevant.

At $k_{\rm min}=1$ we get the following one-loop RG equation

\be \label{8} \frac{d\Delta_1/T^2}{d\ln L}=\big(1-\frac{T}{\pi
J}\big)\frac{\Delta_1}{T^2}-\frac{4\Delta_1^2}{\pi^4 J^4}a, \ee
where $a$ is the ultra-violet spacial cutoff. The fixed point
solution of Eq. (\ref{8}) is $\Delta_1^*=(1-T/(\pi
J))\pi^2J^2/(4a)$. This result is valid, if the dimensionless
combination $\Delta_1a/J^2$ is small.

To obtain the correlation function we add an infinitesimal
contribution $\lambda |\phi^{a=1}_{q=k}|^2/T$ to the replica
Hamiltonian (\ref{7}). The correlation function
$\langle|\phi^1_k|^2\rangle=-T dZ/d\lambda$, where $Z$ is the
replica partition function. To find the contribution to $Z$
proportional to $\lambda$, we calculate $\phi$-independent
corrections to $H_R$ (\ref{7}) generated at each RG step. No
$\lambda$-dependence is present in the first order in $\Delta_1$.
Hence, we have to go to the second order in $\Delta_1$.
This gives us the following result:

\be
\label{9}\langle(\phi^*(x,z=0)-\phi^*(y,z=0))^2\rangle=\frac{2T}{\pi
J}\ln \frac{|x-y|}{a} + \frac{\pi^2}{8}\left( 1-\frac{T}{\pi J}\
\right) ^2\ln \frac{|x-y|}{a} \ee where the first term represents
the effect of thermal fluctuations and the second one represents
the effect of disorder. At $z>0$ the correlation function has the
same structure (\ref{9}), if $|x-y|$ is large enough. At $k_{\rm
min}>1$ the result can be obtained from Eq. (\ref{9}) by the
transformation $\phi^*\rightarrow k\phi^*$, $T\rightarrow k^2 T$.


The correlation function (\ref{9}) is obtained in the vortex-free
model. In the absence of disorder, the
Berezinsky-Kosterlitz-Thouless transition occurs at $T_{\rm
BKT}=\pi J/2$. This temperature is lower than $T_c=\pi J/k_{\rm
min}^2$ in the random-field case $k_{\rm min}=1$. Thus, vortices
lead to a breakdown of our result in the presence of the
random-field disorder. What is the effect of vortices at $k_{\rm
min}>1$? To answer this question we compare the bulk elastic
energy of a vortex in the system of size $L$ and the energy of its
interaction with the surface. The elastic energy $E_{\rm bulk}\sim
J\ln L$. An estimation of the disorder energy must take into
account the renormalization of the disorder strength at large
scales. The fact that we obtain a fixed point $\Delta^*$ for
$\Delta$ means that the effective disorder strength at scale $l$
is of order $\Delta^* a/l$ (rescaling at the RG steps!). The
average square of the disorder energy scales hence as $E^2_{\rm
dis}\sim\int^L dl \Delta^*/l\sim\Delta^* \ln L$. There are $\sim
L^2$ possible positions of the vortex. Assuming a Gaussian
distribution for $E_{\rm dis}$ we get the following probability of
such a disorder realization that the creation of a vortex is
favorable: $p\sim L^2\exp(-J^2 \ln^2 L/(\Delta^* \ln L))$. One
sees that $p\ll 1$, if $J^2\gg\Delta^*$. This justifies our
vortex-free approximation and the result (\ref{9}) at the vicinity
of the critical temperature $T_c$ as the effective disorder
strength $\Delta^*$ is small in that region.

What happens if both surface and bulk disorder is present? This
question can be easily answered in the framework of the Larkin
model \cite{Larkin}. Let the average square of the bulk random
field be $\Delta_{\rm bulk}$. The correlation function in the
presence of bulk disorder only \cite{Larkin} is $\langle
(\phi(x_1)-\phi(x_2))^2\rangle\sim \Delta_{\rm
bulk}|x_1-x_2|/J^2$. This should be compared with the correlation
function in the presence of  surface disorder only (\ref{G}). One
can see that the surface random field is the main source of order
parameter fluctuations at scales
$r=|x_1-x_2|<R_c\sim\Delta/\Delta_{\rm
bulk}\ln(\Delta/(\Delta_{\rm bulk} a))$. At scales $r>R_c$ bulk
disorder dominates.

In conclusion, we demonstrate that even arbitrarily weak surface
disorder destroys long range order in systems of continuous
symmetry at any temperature. Topological order is not destroyed
and quasi-long or super-quasi-long range order emerges. There are
two quasi-long range ordered phases in the random two-dimensional
$XY$ model. The results of the Letter can be relevant for the
characterization of disordered surfaces because the correlation
functions of the system in contact with the surface contain
information about the surface disorder.

We thank  P.H. Kes and V.L. Pokrovsky for a useful discussion.
This work was supported by the US DOE Office of Science under
contract No. W31-109-ENG-38. DEF acknowledges support from RFBR
grant No. 00-02-17763.

\end{document}